\documentclass[%
 reprint,
 superscriptaddress,showkeys,
 amsmath,amssymb,
 aps,
 prd,
 longbibliography
]{revtex4-2}

\usepackage{graphicx}
\usepackage{xcolor}
\usepackage{booktabs}
\usepackage{xspace}
\usepackage{orcidlink}
\usepackage{hyperref}

\hypersetup{
    pdfnewwindow=true,    
    colorlinks=true,      
    linkcolor=blue,       
    citecolor=blue,       
    filecolor=blue,       
    urlcolor=blue         
}

\begin{document}



\title{Complementary Time- and Distance-Based Methods for Cosmogenic \(^{9}\)Li/\(^{8}\)He Background Estimation}



%

\author{Cailian Jiang \orcidlink{0000-0002-3280-0542}}
\email{cailianjiang@smail.nju.edu.cn}
\affiliation{Nanjing University, Nanjing 210093, China}
\affiliation{Institute of High Energy Physics, Chinese Academy of Sciences, Beijing 100049, China}

\author{Qishan Liu \orcidlink{0000-0003-1437-6829}}
\email{liuqs@ihep.ac.cn}
\affiliation{Institute of High Energy Physics, Chinese Academy of Sciences, Beijing 100049, China}
\affiliation{China Center of Advanced Science and Technology, Beijing 100190, China}

\author{Liangjian Wen}
\affiliation{Institute of High Energy Physics, Chinese Academy of Sciences, Beijing 100049, China}

\author{Gaosong Li}
\affiliation{Institute of High Energy Physics, Chinese Academy of Sciences, Beijing 100049, China}

\author{Zeyuan Yu}
\affiliation{Institute of High Energy Physics, Chinese Academy of Sciences, Beijing 100049, China}


\author{Wanlei Guo}
\affiliation{Institute of High Energy Physics, Chinese Academy of Sciences, Beijing 100049, China}

\author{Lei Zhang}
\email{leizhang1801@nju.edu.cn}
\affiliation{Nanjing University, Nanjing 210093, China}

\date{\today}



\begin{abstract}
Cosmogenic $^{9}$Li and $^{8}$He isotopes constitute an important
correlated background in low-energy neutrino experiments because their
$\beta$-delayed neutron decay signatures can mimic inverse beta decay signals.
Conventional estimates based on the time since the last muon become
challenging at high muon rates, while muon-related vetoes further
reduce the residual isotope statistics. We extend the conventional time fit to an event-level muon-categorized joint time (J-MuCAT) fit, which uses the time to the most recent preceding muon in each energy-loss category. A complementary estimate is obtained from the candidate-to-muon-track distance distribution (TraDiTS). The muon-uncorrelated component is determined from the far-distance region and subtracted. This estimate is then used to constrain the J-MuCAT fit, defining the distance-constrained J-MuCAT (DCJ-MuCAT) fit.
In detector-level simulation with successive cosmogenic-background vetoes, DCJ-MuCAT reduces the statistical uncertainty by more than $40\%$ relative to J-MuCAT and by more than $10\%$ relative to the TraDiTS.
The fitted results remain consistent with the simulation truth. Applicability studies further show good performance over a broad range of muon rates and isotope fractions. The proposed framework provides a practical
approach for estimating residual $^{9}$Li/$^{8}$He backgrounds in
large neutrino detectors.
\end{abstract}




\maketitle





\newcommand{\li}{$^{9}$Li\xspace}
\newcommand{\he}{$^{8}$He\xspace}
\newcommand{\lihe}{$^{9}$Li/$^{8}$He\xspace}

\section{Introduction}\label{sec:intro}

Muon-induced radioactive isotopes constitute an important source of
correlated background in underground neutrino
experiments~\cite{DayaBay:2012fng,DoubleChooz:2019qbj,
RENO:2024msr,JUNO:2025gmd}. Cosmic-ray muons traversing a detector can
induce nuclear spallation and secondary-particle cascades, producing
cosmogenic isotopes in the detector. Among them, \li and \he are
particularly important because of their $\beta$-delayed neutron-emission
decays. In these decays, the $\beta$ particle produces a prompt-like
signal, while the emitted neutron subsequently thermalizes and is
captured, forming a prompt-delayed coincidence that closely mimics the
inverse beta decay (IBD) signature. Beyond reactor antineutrino
measurements, cosmogenic isotope backgrounds are also relevant to other
low-energy neutrino studies, including diffuse supernova neutrino
background (DSNB) searches and solar-neutrino
observations~\cite{JUNO:2022lpc,BOREXINO:2018ohr}. Accurate estimation
of the residual \lihe background is therefore essential. In direct dark-matter searches
using liquid argon or liquid xenon, cosmogenic $\beta$-delayed neutrons
can produce single-scatter nuclear recoils that mimic WIMP
signals~\cite{Empl:2014ona,Pec:2023yic}. Measurements of isotope
production yields as a function of muon energy also provide benchmarks
for hadronic interaction models implemented in GEANT4 and
FLUKA~\cite{KamLAND:2009zwo,DoubleChooz:2018kvj}.

Measurements of cosmogenic \lihe production in liquid scintillator have
been reported by several experiments using different approaches to
determine the isotope contribution~\cite{Hagner:2000xb,
KamLAND:2009zwo,Borexino:2013id,DoubleChooz:2018kvj,
RENO:2022xbr,DayaBay:2024xye}. The conventional time-since-last-muon (TSLM) fit is one of the most widely used
approaches~\cite{Wen:2006hx,KamLAND:2009zwo}.
It fits the time distribution of selected candidates relative to
preceding muons and separates the isotope-correlated component from
uncorrelated candidates using their different time distributions.
The method is straightforward to implement and requires only the
temporal information of the candidates and selected muons. The method is statistically effective when the
selected-muon rate is sufficiently low. At higher muon rates, however,
the statistical separation between the isotope and uncorrelated
components deteriorates, reducing the sensitivity of the extraction.
Muon-energy subdivision and cosmogenic-muon enrichment have also been
used to improve the determination of the \lihe contribution. Daya Bay,
for example, divided preceding muons according to their visible energy
and performed time-based estimates for the corresponding muon
samples~\cite{DayaBay:2012yjv}. Double Chooz used the neutron
multiplicity following a muon, together with other muon-related
information, to obtain a sample enriched in cosmogenic
isotopes~\cite{DoubleChooz:2018kvj}. These approaches reduce the
effective muon rate or increase the isotope fraction, improving the
sensitivity of the extraction. However, muon-energy subdivision treats
the corresponding samples separately, so the preceding-muon times from
different categories are not retained simultaneously for each
candidate. Neutron-based enrichment additionally requires a tagging
selection, whose efficiency and uncertainty must be accounted for when
inferring the \lihe contribution. Another approach exploits the prompt-energy spectrum of \lihe. RENO determined the isotope contribution in a high-energy, \lihe-enriched region and extrapolated it to the lower-energy analysis region using the measured isotope spectrum~\cite{RENO:2022xbr}. This approach is less dependent on identifying the parent muon, but it relies on the isotope spectral shape, the treatment of other high-energy backgrounds, and the extrapolation between energy regions.

To improve the time-based extraction at high muon rates, this work
develops an event-level muon-categorized joint time (J-MuCAT) fit.
The selected muons are divided into categories according to their
energy loss in the detector.
For each candidate, the fit retains
the time since the most recent preceding muon in each category, together
with the corresponding muon rate and isotope contribution.
This categorization reduces the muon rate associated with each time
component while preserving differences in isotope production among the
muon categories.
Unlike approaches that treat different muon categories as separate
samples, J-MuCAT uses the timing information from all categories
simultaneously in a common event-level likelihood, avoiding the assignment
of each candidate to a single category and making fuller use of the
available timing information. Together, these features improve the separation of \lihe from
uncorrelated candidates.

Spatial information has also been used in cosmogenic-background
measurements. Double Chooz combined the candidate-to-muon-track
distance with neutron multiplicity to select a cosmogenic-enriched
sample~\cite{DoubleChooz:2018kvj}, while Borexino characterized the
spatial distributions of muon-induced products relative to reconstructed
muon tracks~\cite{Borexino:2013id}. Such spatial information provides
strong discrimination, but has mainly been used for event selection,
enrichment, or characterization rather than to directly determine the
\lihe contribution from the distance distribution.
Muon-related vetoes further suppress cosmogenic backgrounds but reduce
the remaining isotope statistics. Consequently, the time-based
extraction can become weakly constrained after strong vetoes, even with
the additional category-dependent information.
To obtain an independent spatial estimate, this work introduces the
track-distance template subtraction (TraDiTS) method. The uncorrelated
distance component is determined from a far-distance region and
extrapolated to the near-distance region, where the excess gives the
isotope contribution without requiring an explicit \lihe distance
template. This estimate is then used to constrain the J-MuCAT fit,
forming the distance-constrained J-MuCAT (DCJ-MuCAT) fit. The approach
combines temporal and spatial information without requiring a
multidimensional isotope model and improves the extraction when the
residual isotope statistics are limited.

This work presents a methodological study of time- and distance-based
approaches for estimating residual \lihe backgrounds. The performance
of the conventional TSLM fit, J-MuCAT fit, TraDiTS method, and
DCJ-MuCAT fit is evaluated with detector-level simulation under
successive muon-related vetoes, and their applicability under different muon rates, isotope fractions, and sample statistics is further investigated.
The main analysis dependences, including muon categorization,
muon-association windows, track-distance template configuration, and
assumed isotope composition, are investigated. Statistical and
reconstruction-related robustness checks are also performed,
particularly under the limited isotope statistics remaining after the
vetoes.

\section{Methods}
\subsection{Time-Based Method}
\label{sec:time_method}

\subsubsection{Conventional Time-Since-Last-Muon Fit}
\label{subsec:conventional_time}

The conventional time-since-last-muon (TSLM) method uses the time since the last selected muon. The selected muons are
modeled as a stationary Poisson process with rate $R_\mu$. For a candidate
uncorrelated with the muon sequence, the time $t$ since the last selected
muon follows
\begin{equation}
    f_{\rm uncorr}(t)
    =
    R_\mu e^{-R_\mu t},
    \qquad t\geq 0.
    \label{eq:uncorrelated_time_pdf}
\end{equation}

For isotope $i$, the parent muon is not necessarily the last selected muon
before the decay because additional selected muons may occur in between.
The resulting TSLM distribution is
\begin{equation}
    f_i(t)
    =
    \lambda_i e^{-\lambda_i t},
    \qquad
    \lambda_i
    =
    \frac{1}{\tau_i}+R_\mu,
    \label{eq:isotope_time_pdf}
\end{equation}
where $\tau_i$ is the mean lifetime of isotope $i$. The $R_\mu$ term
accounts for intervening selected muons that may become the last muon
before the isotope decay.

In the low-muon-rate limit, $R_\mu\ll\tau_i^{-1}$, the isotope component
approaches its intrinsic decay-time distribution. The uncorrelated
component is approximately constant when the fit interval is much shorter
than the mean time between selected muons.

For a finite fit interval $[t_{\min},t_{\max}]$, each component is
normalized within that interval:
\begin{equation}
    p_x(t)
    =
    \frac{f_x(t)}
    {\displaystyle
     \int_{t_{\min}}^{t_{\max}} f_x(t')\,dt'},
    \label{eq:range_normalization}
\end{equation}
where $x$ denotes either the isotope or uncorrelated component.

The observed TSLM distribution contains isotope-correlated
and uncorrelated components. For a sample containing $N_{\rm obs}$
candidates, the normalized mixture PDF is
\begin{equation}
    q(t)
    =
    \eta_{\rm iso}p_{\rm iso}(t)
    +
    \left(1-\eta_{\rm iso}\right)p_{\rm uncorr}(t),
    \label{eq:conventional_mixture_pdf}
\end{equation}
where $\eta_{\rm iso}=N_{\rm iso}/N_{\rm obs}$ is the isotope fraction and
$N_{\rm iso}$ is the fitted isotope count. The unbinned likelihood is
\begin{equation}
    \mathcal{L}_{\rm conv}
    =
    \prod_{m=1}^{N_{\rm obs}}q(t_m).
    \label{eq:conventional_unbinned_likelihood}
\end{equation}

The conventional formulation treats all selected muons as a single
Poisson process. At high muon rates, intervening muons weaken the
correlation with the parent muon. Dividing the muon sample into categories
reduces the muon rate within each category while retaining differences in
isotope production. This motivates the extended time-based method
described below.

\subsubsection{Muon-Categorized Joint Time Fit}
\label{subsec:category_time}

To improve the applicability of the time-based method at high muon rates, the selected muons are divided into $K$ disjoint
categories according to a chosen muon observable. In this study, the
categorization is based on the muon energy loss in the target medium.
For each candidate $m$, $t_{k,m}$ denotes the time elapsed since the last
muon in category $k$, and the candidate is represented by
\begin{equation}
    \mathbf{t}_m
    =
    \left(t_{1,m},t_{2,m},\ldots,t_{K,m}\right).
    \label{eq:category_time_vector}
\end{equation}

The muon rate in category $k$ is denoted by $R_{\mu,k}$. Within the fit
interval $[t_{\min},t_{\max}]$, the normalized time PDF for an
uncorrelated candidate is
\begin{equation}
    u_k(t)
    =
    \frac{
        R_{\mu,k}e^{-R_{\mu,k}t}
    }{
        e^{-R_{\mu,k}t_{\min}}
        -
        e^{-R_{\mu,k}t_{\max}}
    }.
    \label{eq:category_uncorr_pdf}
\end{equation}

For isotope $i$ produced by a muon in category $k$, subsequent muons
in the same category may occur before the isotope decays, as in
Eq.~\eqref{eq:isotope_time_pdf}. This gives an effective slope
$\alpha_{i,k} = 1/\tau_i + R_{\mu,k}$ and the normalized PDF
\begin{equation}
    s_{i,k}(t)
    =
    \frac{
        \alpha_{i,k}e^{-\alpha_{i,k}t}
    }{
        e^{-\alpha_{i,k}t_{\min}}
        -
        e^{-\alpha_{i,k}t_{\max}}
    }.
    \label{eq:category_isotope_pdf}
\end{equation}

For an uncorrelated candidate, the times to muons in all categories follow
the corresponding uncorrelated PDFs:
\begin{equation}
    U(\mathbf{t}_m)
    =
    \prod_{j=1}^{K}
    u_j(t_{j,m}).
    \label{eq:joint_uncorr_pdf}
\end{equation}

If an isotope is associated with a muon in category $k$, $t_{k,m}$
follows the isotope-correlated distribution, while the times to muons
in the other categories remain uncorrelated. Including $^{9}$Li and
$^{8}$He with fixed lifetimes and a fixed relative normalization, the
joint isotope PDF is
\begin{equation}
    S_k(\mathbf{t}_m)
    =
    \frac{
        s_{^{9}{\rm Li},k}(t_{k,m})
        +
        r_{\rm He/Li}
        s_{^{8}{\rm He},k}(t_{k,m})
    }{
        1+r_{\rm He/Li}
    }
    \prod_{\substack{j=1\\j\neq k}}^{K}
    u_j(t_{j,m}),
    \label{eq:joint_isotope_pdf}
\end{equation}
where
$r_{\rm He/Li}=N(^{8}\mathrm{He})/N(^{9}\mathrm{Li})$
denotes the relative contribution of the selected $^{8}\mathrm{He}$ and
$^{9}\mathrm{Li}$ events and is fixed to $0.047$ in the nominal fit.

Each candidate enters the extended unbinned likelihood once, with all
category-dependent times treated jointly:
\begin{equation}
\begin{aligned}
    \mathcal{L}_{\rm time}
    ={}&
    \exp\left[
        -N_{\rm uncorr}
        -
        \sum_{k=1}^{K}N_{{\rm iso},k}
    \right]
    \\
    &\times
    \prod_{m=1}^{N_{\rm obs}}
    \Biggl[
        N_{\rm uncorr}U(\mathbf{t}_m)
    \\
    &\quad+
        \sum_{k=1}^{K}
        N_{{\rm iso},k}S_k(\mathbf{t}_m)
    \Biggr],
\end{aligned}
\label{eq:joint_time_likelihood}
\end{equation}
where $N_{{\rm iso},k}$ is the total $^{9}$Li and $^{8}$He contribution
associated with muon category $k$. The total isotope contribution is
\begin{equation}
    N_{\rm iso}
    =
    \sum_{k=1}^{K}N_{{\rm iso},k}.
    \label{eq:category_total_count}
\end{equation}

The category-specific muon rates $R_{\mu,k}$ are free
parameters in the joint fit. Their fitted values are compared with the
rates obtained directly from the selected muon sample as a consistency
check of the time model. The impact of additional muon-rate information is examined
separately in Sec.~\ref{sec:time_applicability}.
\subsection{Track-Distance Template Subtraction Method}
\label{sec:distance_method}
The track-distance template subtraction (TraDiTS) method exploits the spatial correlation
between cosmogenic isotopes and muon tracks. For each candidate--muon pair within a given time window, $d$ denotes the candidate-to-muon-track distance. Isotope candidates are expected to
concentrate at small $d$, whereas muon-uncorrelated candidates follow a
distribution determined mainly by the detector geometry and the muon-track
distribution. Assuming that the isotope contribution is negligible beyond a suitably chosen distance, the uncorrelated component is determined from the far-distance region. The isotope contribution is then obtained from the excess in the near-track region.

An uncorrelated distance template, denoted by $T_{\rm uncorr}(d)$, is constructed to describe the candidate-to-muon-track distance distribution of muon-uncorrelated events. Let $\Omega_{\rm near}$ and $\Omega_{\rm far}$ denote the near-track and far-distance regions, respectively. The template is fitted to the candidate-muon pair distribution in $\Omega_{\rm far}$, where the isotope contribution is assumed to be negligible:
\begin{equation}
    F_{\rm far}(d)
    =
    \alpha T_{\rm uncorr}(d),
    \qquad d \in \Omega_{\rm far},
    \label{eq:distance_far_fit}
\end{equation}
where $\alpha$ is a free normalization parameter.

The fitted value $\hat{\alpha}$ is used to estimate the uncorrelated
contribution in the near-track region:
\begin{equation}
    \hat{N}_{\rm uncorr}(\Omega_{\rm near})
    =
    \hat{\alpha}
    \int_{\Omega_{\rm near}}
    T_{\rm uncorr}(d)\,{\rm d}d.
    \label{eq:distance_uncorr_near}
\end{equation}
The isotope contribution is then obtained as
\begin{equation}
    \hat{N}_{\rm iso}^{\rm dist}
    =
    N_{\rm obs}(\Omega_{\rm near})
    -
    \hat{N}_{\rm uncorr}(\Omega_{\rm near}).
    \label{eq:distance_near_residual}
\end{equation}

Under this assumption, Eq.~\eqref{eq:distance_near_residual} provides the
total isotope contribution without requiring an explicit isotope distance
template. The dependence on the distance fit range and histogram binning is evaluated in the systematic uncertainty study.

\subsection{Distance-Constrained J-MuCAT Fit}
\label{sec:combined_method}

The distance-template result is incorporated into the
J-MuCAT fit as a Gaussian constraint on the total
isotope contribution. Let
$\hat{N}_{\rm iso}^{\rm dist}$ and $\sigma_{\rm dist}^{\rm stat}$
denote the isotope contribution obtained with the distance method
and its statistical uncertainty, respectively. The constrained
likelihood is defined as
\begin{equation}
    \mathcal{L}_{\rm comb}
    =
    \mathcal{L}_{\rm time}
    \exp\left[
        -\frac{
            \left(
                N_{\rm iso}
                -
                \hat{N}_{\rm iso}^{\rm dist}
            \right)^2
        }{
            2\left(\sigma_{\rm dist}^{\rm stat}\right)^2
        }
    \right],
    \label{eq:combined_likelihood}
\end{equation}
Only the statistical uncertainty of the distance result is included in the
Gaussian constraint. Distance-related systematic effects are
evaluated separately by repeating the complete analysis under the
corresponding variations. The fitted isotope contribution and its nominal statistical uncertainty are obtained from the profile likelihood based on
$\mathcal{L}_{\rm comb}$.

The time and distance estimates are obtained from the same candidate
sample and are therefore statistically correlated. The Gaussian term is
thus not treated as an independent external likelihood. This correlation
is evaluated using a paired bootstrap study, in which both estimates are
recalculated from the same resampled candidate set. The results are
discussed in Sec.~\ref{sec:robustness_checks}.

A two-dimensional event-level fit in $(t,d)$ is not adopted in this
work. Such a fit would require a joint model of the time and distance
distributions, including an explicit isotope-distance distribution that
is difficult to determine reliably. The limited isotope statistics
remaining after the vetoes would further weaken the constraints on such
a model. A factorized treatment using separate one-dimensional PDFs
would instead introduce an additional assumption on the dependence
between $t$ and $d$. For these reasons, the spatial information is
incorporated through the distance-constrained time fit rather than an
event-level two-dimensional fit.

\section{Simulation setup}
\label{sec:simulation_setup}

A simulation sample is constructed to validate the proposed methods under
controlled conditions. It consists of simulated muons and their associated
cosmogenic isotopes, together with independently generated IBD events. The muons and associated cosmogenic isotopes are simulated using a standalone Geant4 simulation (version 11.0.3) of a representative large liquid-scintillator detector based on the JUNO design~\cite{JUNO:2021vlw}. The geometry includes a spherical liquid-scintillator target with a diameter of $35.4~\mathrm{m}$, enclosed by an acrylic vessel. It also includes the surrounding water region, experimental hall, and rock overburden. The target material is modeled as a standard liquid scintillator. Optical-photon production and transport are disabled because a detailed optical detector response is not required for the present method validation. For the present method-validation sample, only events with a single muon traversing the liquid-scintillator target are retained, giving an effective selected-muon rate of approximately 2.5 Hz. Independently generated IBD events are added at a rate of approximately $2650~\mathrm{day^{-1}}$, corresponding to about $47$ times the reference rate of $57~\mathrm{day^{-1}}$~\cite{JUNO:2024jaw}. This enhanced rate increases the uncorrelated-event statistics for the method-validation study.

\begin{figure}
    \centering
    \includegraphics[width=1\columnwidth]{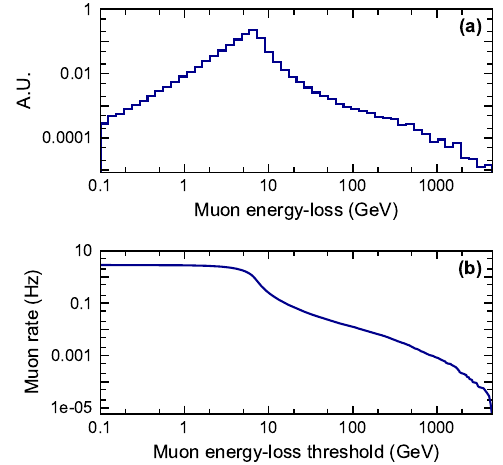}
    \caption{
Muon energy-loss distribution and cumulative muon rate in the liquid
scintillator (LS). (a) Muon energy-loss distribution. (b) Muon rate
above a given energy-loss threshold.
}
    \label{fig:muonenergy}
\end{figure}

Muon propagation and cosmogenic-isotope production are simulated using a nominal physics configuration. It includes electromagnetic and hadronic interactions, particle and radioactive decays, and muon stopping. 
Cosmic muons are
generated at the surrounding rock boundary and propagated through the rock, water region, and detector volumes. For each muon, its trajectory and the associated isotope, neutron-capture, and decay information are recorded. The recorded quantities include positions, energies, and times relative to the parent muon. Fig.~\ref{fig:muonenergy} shows the muon energy-loss distribution in the liquid scintillator and the corresponding cumulative muon rate.

To construct the analysis sample, absolute times are assigned to the selected muons. The associated \lihe decay events are then placed according to their simulated times relative to the parent muons. Independently generated IBD events are inserted according to the chosen IBD rate. All events are then ordered by absolute time to form a single event sequence. 

\section{Results}
\begin{figure}[t]
    \centering
    \includegraphics[width=1\columnwidth]{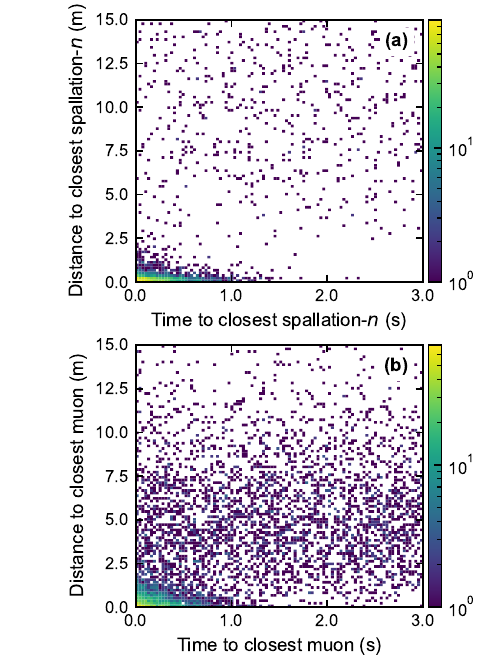}
    \caption{
Time and distance correlations of selected candidates after the basic
event selection. (a) Time and distance relative to the closest
spallation neutron. (b) Time and distance relative to the closest muon
track.
}
    \label{fig:2dmuontrack}
\end{figure}

We evaluate the proposed methods using the simulation sample under three progressively stricter selections. These are the basic IBD event selection, the spallation-neutron veto, and the combined spallation-neutron and muon-track vetoes. These selections are based on the criteria used in
Ref.~\cite{JUNO:2025gmd}, and the full selection requirements are
summarized in Table~\ref{tab:eventsel}. 
At each selection stage, the different methods are applied to the same candidate sample for direct comparison. The
nominal configuration of each method is specified in the corresponding
subsection. The simulation-truth \lihe event counts at each selection stage are listed in Table~\ref{tab:time_fit_results} and used as the reference for the comparisons below. Fig.~\ref{fig:2dmuontrack} shows the time and distance correlations of the selected candidates with spallation neutrons and muon tracks after the basic selection. A pronounced concentration of events is observed at short times and distances, corresponding to a \lihe-enriched region. The remaining events are dominated by muon-uncorrelated IBD events. These correlations form the basis of the spallation-neutron and muon-track vetoes commonly used in reactor-antineutrino analyses
~\cite{DayaBay:2012yjv,JUNO:2025gmd}.


\begin{table}[!t]
    \caption{Summary of selection criteria.}
    \label{tab:eventsel}
    \centering
    \footnotesize
    \setlength{\tabcolsep}{3pt}
    \renewcommand{\arraystretch}{1.2}

    \begin{tabular}{
        p{0.35\columnwidth}
        p{0.60\columnwidth}
    }
        \toprule
        Quantity & Value \\
        \midrule

        Prompt energy
        & $[0.7, 12]\,\mathrm{MeV}$ \\

        Delayed energy
        & $[2.0, 2.5]\,\mathrm{MeV}$ and
          $[4.4, 5.5]\,\mathrm{MeV}$ \\

        Coincidence time
        & $[5, 1000]\,\mu\mathrm{s}$ \\

        Relative distance
        & $< 1.5\,\mathrm{m}$ \\

        Spallation neutron veto
        & Reject candidates with
          $\Delta r \leq 4\,\mathrm{m}$ and
          $\Delta t \leq 1.2\,\mathrm{s}$
          relative to a spallation neutron \\

        Muon track veto
        & Reject candidates with
          $\Delta r \leq 2.5\,\mathrm{m}$ and
          $\Delta t \leq 0.5\,\mathrm{s}$
          relative to a muon track \\
        \bottomrule
    \end{tabular}
\end{table}

    
    
    

\subsection{Muon-Categorized Joint Time Fit}
As described in Sec.~\ref{sec:time_method}, the time-based method
uses an event-level joint likelihood based on the times to the most recent
preceding muon in each category. In the nominal configuration, the
preceding muons are divided into four categories according to their energy
loss in the liquid scintillator:
$0.1\leq E_{\mu}<4.4~\mathrm{GeV}$,
$4.4\leq E_{\mu}<6.3~\mathrm{GeV}$,
$6.3\leq E_{\mu}<7.5~\mathrm{GeV}$, and
$E_{\mu}\geq7.5~\mathrm{GeV}$.
An energy-loss threshold of $0.1~\mathrm{GeV}$ is used to identify muons
traversing the liquid scintillator. The category boundaries are not optimized in this work. The simulation study is intended to validate the method rather than determine an optimal muon-energy partition.

\begin{figure*}[t]
    \centering
    \includegraphics[width=1\textwidth]
    {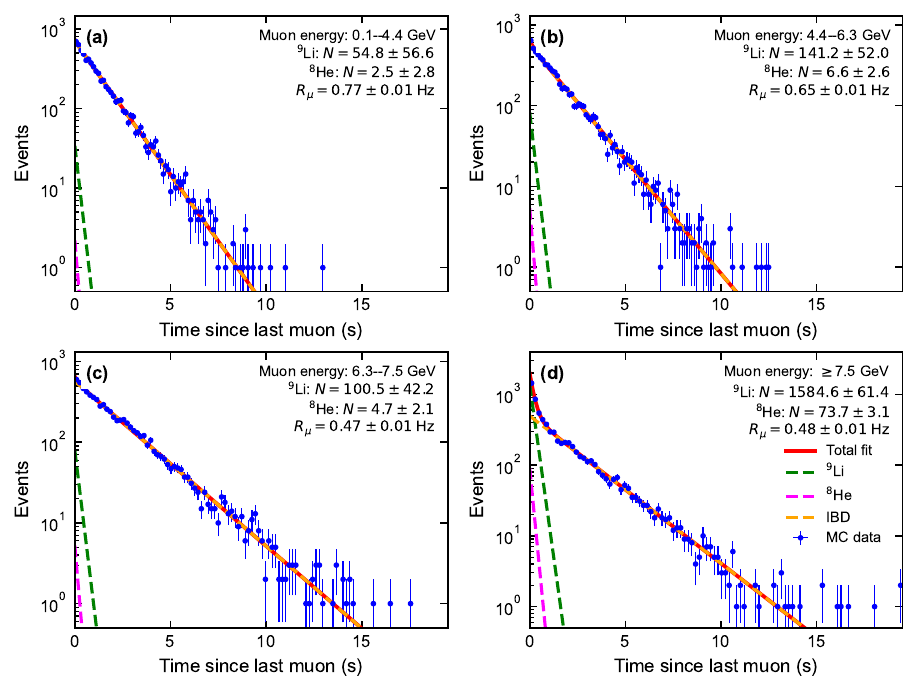}
    \caption{
J-MuCAT fit results after the basic event selection.
The four panels correspond to muon energy-loss intervals of
$0.1$--$4.4~\mathrm{GeV}$,
$4.4$--$6.3~\mathrm{GeV}$,
$6.3$--$7.5~\mathrm{GeV}$, and
$\geq 7.5~\mathrm{GeV}$.
The blue points show the simulation sample, while the red curves show the total fitted model.
The orange, green, and purple curves represent the muon-uncorrelated IBD,
$^{9}$Li, and $^{8}$He components, respectively.
}
    \label{fig:timefitprimary}
\end{figure*}

The nominal preceding-muon search window is $20~\mathrm{s}$. The event-level joint likelihood in Eq.~\eqref{eq:joint_time_likelihood} is constructed from the times to the most recent preceding muon in each category within this window. The $^{9}$Li and $^{8}$He mean lifetimes are fixed in the fit at 257~ms and 172~ms, respectively~\cite{nndc}. The relative isotope composition, $N(^{8}\mathrm{He})/N(^{9}\mathrm{Li})$, is fixed to $0.047$ in the nominal analysis, as determined from the MC truth. For comparison, Ref.~\cite{DayaBay:2024xye} reports a ratio of $0.0143$, which is included as an alternative value in the systematic study. The category-specific isotope contributions and muon rates are determined by the fit. The fitted distributions after the basic event selection are shown in Fig.~\ref{fig:timefitprimary}. The fitted model describes the simulation sample over the selected time range.  The fitted
\lihe contributions in the four muon energy-loss categories are
$57.3\pm59.2$, $147.8\pm54.4$, $105.2\pm44.1$, and
$1658.3\pm64.2$ events, respectively, compared with the corresponding
simulation-truth values of 86, 115, 107, and 1673 events. All four fitted
contributions are consistent with the simulation truth within their
statistical uncertainties. In particular, the highest-energy category
accounts for approximately $84.2\%$ of the fitted isotope contribution,
consistent with the simulation-truth fraction of $84.5\%$. This agreement provides a closure test of the category decomposition. The corresponding category-specific muon rates determined directly from the selected muon sample are
$R_{\mu,k}^{\mathrm{MC}}
=(0.77,\,0.66,\,
0.47,\,0.48)\,\mathrm{Hz}$,
in the category order defined above. The fitted rates in Fig.~\ref{fig:timefitprimary} are consistent with these directly determined values within the fit uncertainties.

The fit is repeated after the basic event selection, the spallation-neutron veto, and the combined spallation-neutron and muon-track vetoes. For comparison, the conventional time fit is applied to the same candidate samples using the same preceding-muon search window, isotope lifetimes, and fixed isotope composition. The fitted isotope contributions from both time fits are summarized in Table~\ref{tab:time_fit_results}. Compared with the conventional TSLM fit, the J-MuCAT fit gives smaller statistical uncertainties at all three selection stages. After the combined veto, the conventional fit
gives a small positive best-fit value that lies well below the
simulation truth and close to the physical boundary at zero. In contrast, the J-MuCAT fit gives a finite isotope contribution closer to the simulation-truth value, with a smaller statistical uncertainty. The residual contribution nevertheless remains weakly constrained because
of the limited isotope statistics.






\begin{table*}[htbp]
    \centering
    \caption{Comparison of the \lihe event contributions obtained with the
    conventional TSLM and J-MuCAT fits. The simulation-truth values are
    also shown. The uncertainties are statistical only.}
    \label{tab:time_fit_results}
    \begin{tabular}{lccc}
        \hline
        Selection & Simulation truth & TSLM & J-MuCAT \\
        \hline
        Basic event selection
        & 1981
        & $2065.1 \pm 177.2$
        & $1968.7 \pm 110.8$ \\
        After SPN veto
        & 114
        & $144.7 \pm 125.9$
        & $123.9 \pm 85.0$ \\
        After SPN + muon-track vetoes
        & 33
        & $4.2 \pm 121.5$
        & $31.3 \pm 75.8$ \\
        \hline
    \end{tabular}
\end{table*}
\subsection{Track-Distance Template Subtraction Method}

The performance of the TraDiTS method is
evaluated using the same simulation sample and event selections as the
time-correlation method.
The nominal configuration uses a $3~\mathrm{s}$ preceding-muon search window and 100 histogram bins over the far-distance fit range of $5$--$25~\mathrm{m}$. Each candidate is
paired with all preceding muon tracks within the search window.
As described in Sec.~\ref{sec:distance_method}, the muon-uncorrelated
component is modeled using a distance template constructed from uniformly
distributed IBD events and fitted to the candidate--muon pair distribution
in the far-distance region. The cosmogenic-isotope contribution is then obtained from the residual in the near-track region after subtraction of the fitted template.

\begin{figure}[t]
    \centering
    \includegraphics[width=\columnwidth]{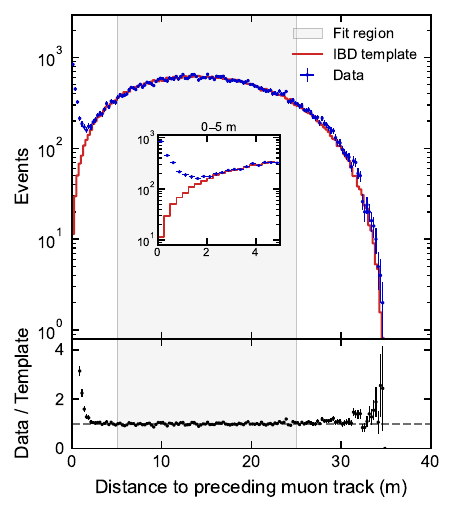}
    \caption{
TraDiTS result after the basic event selection.
The upper panel shows the candidate-to-muon-track distance
distribution in the simulation sample together with the fitted
IBD distance template.
The shaded area indicates the far-distance region used in the fit.
The lower panel shows the ratio of the simulation sample to the fitted template.
The template describes the distribution in the fit region, while the excess
at small distances corresponds to the cosmogenic-isotope component.
}
    \label{fig:distancefitprimary}
\end{figure}

The distance distribution after the basic event selection is shown in Fig.~\ref{fig:distancefitprimary}. The fitted template describes the candidate--muon pair distribution in the far-distance fit region. At small distances, a clear excess over the fitted template is observed, corresponding to the cosmogenic-isotope component.

As shown in Fig.~\ref{fig:nominal_comparison}, the extracted isotope
contribution decreases after each successive veto, consistent with the
expected suppression of cosmogenic backgrounds. Although the residual
estimate becomes less precise after the vetoes, its statistical
uncertainties remain smaller than those from the J-MuCAT fit under the
same selections. The far-distance region continues to constrain the
muon-uncorrelated component even when the residual isotope statistics
are limited.

    
    
    

\subsection{Distance-Constrained J-MuCAT Fit}

The final \lihe event contribution is obtained with the distance-constrained J-MuCAT (DCJ-MuCAT) fit described in Sec.~\ref{sec:combined_method}. The nominal fit uses the
four-category time configuration and the nominal track-distance estimate defined above.
The distance estimate and its statistical uncertainty are used as a Gaussian constraint on the total isotope contribution in the J-MuCAT fit.

\begin{figure*}[htbp]
    \centering
    \includegraphics[width=1\textwidth]{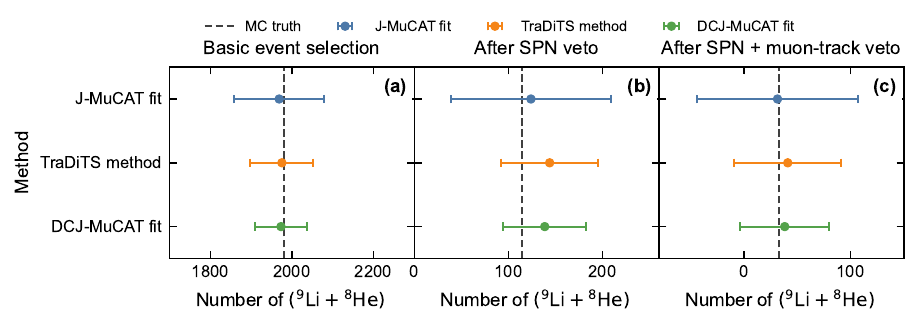}
    \caption{
Comparison of the fitted \lihe event contributions obtained with the
J-MuCAT fit, TraDiTS method, and DCJ-MuCAT fit.
The three panels correspond to the basic selection, the SPN-veto,
and the combined SPN+track-vetoes, respectively.
The error bars represent statistical uncertainties, and the vertical dashed
lines indicate the corresponding simulation-truth \lihe values.
}
    \label{fig:nominal_comparison}
\end{figure*}

Figure~\ref{fig:nominal_comparison} compares the J-MuCAT fit,
track-distance estimate, and DCJ-MuCAT fit with the true
\lihe contribution in the simulation. For the basic selection, both the time and distance estimates are consistent with the simulation truth within their statistical uncertainties. The constrained fit is also consistent with the simulation truth and has a smaller statistical uncertainty than either individual estimate.
After the spallation-neutron veto, the two individual estimates remain
compatible with each other and with the simulation truth. After the combined
spallation-neutron and muon-track vetoes, the reduced isotope statistics
weaken the time-only extraction, while the distance constraint improves the
precision. For all three event selections, the distance-constrained results
are consistent with the corresponding simulation-truth values. Their
statistical uncertainties are reduced by approximately $43$--$48\%$ relative
to the J-MuCAT fit and by approximately $15$--$18\%$ relative
to the track-distance estimate. The effect of the shared candidate sample on the constrained fit is further evaluated with the paired-bootstrap study in Sec.~\ref{sec:robustness_checks}.

\subsection{Applicability Studies}
\label{sec:time_applicability}

To assess the applicability beyond the nominal configuration, the
DCJ-MuCAT fit is repeated with 0.5 and 2 times the nominal candidate
statistics and with \li-to-IBD ratios from 0.5 to 2 times the nominal
value. In all cases, the fitted contribution remains consistent with
the event-level simulation truth within the statistical uncertainty.
The J-MuCAT and TraDiTS methods are further tested with parametric toys
using total selected-muon rates of 0.1, 0.3, 1, 3, 5, 10, and 30~Hz
and \li-to-IBD ratios of 0.01, 0.03, 0.1, and 1. Each point contains
1000 toys with $N_{\rm IBD}=10^{6}$. The relative bias and spread are
defined as $\overline{(F-T)}/\overline{T}$ and
$\mathrm{std}(F-T)/\overline{T}$, respectively, where $F$ and $T$
denote the fitted and generated isotope contributions in each toy.

Both methods show good performance over a broad region of the tested
parameter space. The TraDiTS estimate remains nearly unbiased over the
full scan, with the relative bias below 0.6\%. Increasing the muon rate
mainly increases the statistical spread; for the lowest \li-to-IBD
ratio of 0.01, the spread increases from 1.9\% at 0.1~Hz to 29.0\%
at 30~Hz. For the J-MuCAT fit, with the category-specific muon rates
and isotope contributions left free, the relative bias remains below
2.0\% and the spread below 10.0\% for \li-to-IBD ratios of 0.03 or
larger and muon rates up to 5~Hz. The accessible rate range increases
with the isotope fraction: at a ratio of 0.03, the bias and spread are
3.1\% and 15.1\% at 10~Hz, while at a ratio of 0.1 they are 5.7\%
and 19.7\% at 30~Hz. Overall, both methods retain good performance
over a wide range of muon rates and isotope fractions, with reduced
precision mainly in the low-isotope-fraction and high-rate region.

The muon-rate dependence of the J-MuCAT fit can be understood through
the category-specific quantity $R_{\mu,k}\tau_{\rm Li}$. For
$^{9}$Li, $\tau_{\rm Li}\simeq0.257$~s, and the temporal association
becomes increasingly ambiguous as $R_{\mu,k}\tau_{\rm Li}$ approaches
unity. Muon categorization lowers $R_{\mu,k}$ and extends the effective
muon-rate range.
Since the present scan uses four muon categories, the observed
high-rate behavior does not represent a universal rate limit; a finer
categorization may further reduce the category-specific rates when
sufficient isotope statistics are available. The scan is also repeated
with the category-specific muon rates fixed to their generated values.
At 30~Hz, the relative bias remains below 5.0\% and the spread below
16.6\% for \li-to-IBD ratios of 0.03 or larger. This shows that
muon-rate information can reduce rate--yield correlations and further
extend the useful high-rate region.

The use of measured muon-rate information is further tested with the
detector-level sample. The category-specific muon rates are measured
using the same muon categories as in the timing fit and included as
Gaussian constraints,
\begin{equation}
\mathcal{L}_{\rm rate}
=
\mathcal{L}_{\rm time}
\prod_{k=1}^{K}
\exp\left[
-\frac{
\left(R_{\mu,k}-\hat R_{\mu,k}\right)^2
}{
2\sigma_{R,k}^{2}
}
\right],
\label{eq:rate_constraint}
\end{equation}
where $\hat R_{\mu,k}$ and $\sigma_{R,k}$ denote the measured rate
and the corresponding constraint width. Motivated by the observed
temporal variations of underground muon rates reported by previous
experiments~\cite{JUNO:2025fpc,Bellini:2012te,MINOS:2009njg}, three constraint widths are considered: the
statistical uncertainty alone, and two broader cases with additional
relative rate uncertainties of 1\% and 2\%. Relative to the free-rate
J-MuCAT fit, the statistical-only constraint reduces the fitted
uncertainty by 10.0\%, 12.1\%, and 32.7\% for the basic, SPN-veto,
and combined-veto selections, respectively. The corresponding
reductions in the fitted statistical uncertainty are 6.8\%, 9.1\%, and 30.3\% for the 1\% case, and
3.5\%, 5.2\%, and 6.6\% for the 2\% case. The fitted central values
vary only modestly among these constraint choices and remain
consistent with the simulation truth within the statistical
uncertainties. For the basic selection, the paired-bootstrap study shows improved fit stability with the statistical-only rate constraint, while the relative bias remains essentially unchanged. Muon-rate information provides an important additional constraint on
the timing fit and extends its useful range. Overall, the methods
remain effective over a broad range of candidate statistics, isotope
fractions, and muon rates under different detector conditions.

\section{Systematic Uncertainties}
\label{sec:systematics}
\subsection{Evaluation Procedure}
The systematic study evaluates the total isotope contribution, $N_{\mathrm{iso}}\equiv
N(^{9}\mathrm{Li}+{}^{8}\mathrm{He})$, obtained from the DCJ-MuCAT fit.

The nominal analysis configuration is used as the reference. For each
systematic source, the corresponding analysis setting is varied while all
other settings are kept fixed. The complete analysis chain is repeated for each variation. This includes the J-MuCAT fit, the TraDiTS method, and the final DCJ-MuCAT fit. The variations are evaluated separately at all three selection stages.

For each systematic source $j$, the uncertainty is defined as the largest
absolute deviation of the distance-constrained result from the nominal
result among the predefined variations,
\begin{equation}
    \Delta_j
    =
    \max_k
    \left|
        N_{j,k}-N_{\mathrm{nom}}
    \right|,
    \label{eq:systematic_source}
\end{equation}
where $k$ labels the variations associated with source $j$.
$N_{j,k}$ denotes the result obtained with variation $k$,
and $N_{\mathrm{nom}}$ is the nominal result at the same selection stage.
The corresponding relative uncertainty is
\begin{equation}
    \delta_j
    =
    \frac{\Delta_j}{N_{\mathrm{nom}}}
    \times 100\%.
    \label{eq:relative_systematic}
\end{equation}

Treating the individual systematic sources as
independent, the total relative systematic uncertainty is calculated as
\begin{equation}
    \delta_{\mathrm{tot}}
    =
    \sqrt{\sum_j \delta_j^2}.
    \label{eq:total_systematic}
\end{equation}

Bootstrap studies are used to evaluate statistical fluctuations and fit
stability and are not included in the systematic uncertainty budget.
The nominal configuration is adopted as a representative working point rather than an extensively optimized configuration. The quoted uncertainties therefore characterize the dependence of the result on alternative analysis settings, rather than symmetric variations around the nominal setting.

\subsection{Sources of Systematic Uncertainty}

The following sources are included in the adopted systematic uncertainty
budget.

\begin{itemize}
    \item \textbf{Muon energy-loss category boundaries.}
To evaluate the dependence on the adopted categorization, one internal
boundary is varied at a time while the remaining boundaries and the number
of categories are kept fixed. The nominal and varied category definitions
included in the scan are summarized in Table~\ref{tab:muon_category_scan}.

    
    
    
    
    
    
    
    
    
\begin{table}[!t]
    \caption{Muon energy-loss category definitions used in the
    category-boundary scan. One internal boundary is varied at a time
    relative to the nominal configuration.}
    \label{tab:muon_category_scan}
    
    \begin{tabular*}{\columnwidth}{
        @{}
        l
        @{\hspace{2.2em}}
        l
        @{\extracolsep{\fill}}
    }
    \toprule
    Boundary setting & Muon energy-loss categories (GeV) \\
    \midrule
    
    Nominal
    & $0.1$--$4.4$, $4.4$--$6.3$, $6.3$--$7.5$, $\geq 7.5$ \\
    
    $4.4 \rightarrow 3.1$
    & $0.1$--$3.1$, $3.1$--$6.3$, $6.3$--$7.5$, $\geq 7.5$ \\
    
    $4.4 \rightarrow 3.8$
    & $0.1$--$3.8$, $3.8$--$6.3$, $6.3$--$7.5$, $\geq 7.5$ \\
    
    $4.4 \rightarrow 5.0$
    & $0.1$--$5.0$, $5.0$--$6.3$, $6.3$--$7.5$, $\geq 7.5$ \\
    
    $6.3 \rightarrow 5.6$
    & $0.1$--$4.4$, $4.4$--$5.6$, $5.6$--$7.5$, $\geq 7.5$ \\
    
    $6.3 \rightarrow 6.9$
    & $0.1$--$4.4$, $4.4$--$6.9$, $6.9$--$7.5$, $\geq 7.5$ \\
    
    $7.5 \rightarrow 9.4$
    & $0.1$--$4.4$, $4.4$--$6.3$, $6.3$--$9.4$, $\geq 9.4$ \\
    
    \bottomrule
    \end{tabular*}
\end{table}
\item \textbf{Muon-association time windows.}
The nominal preceding muon search windows are $20~\mathrm{s}$ for the
J-MuCAT fit and $3~\mathrm{s}$ for the
TraDiTS method. The
time-fit window is varied among 10, 15, and
$20~\mathrm{s}$, while the distance-method window is varied among 2, 3,
and $4~\mathrm{s}$. Each window is varied independently, with the other
window fixed at its nominal value. Varying the time-fit window can change the accepted candidate sample by modifying the availability of preceding muons in the required categories. In contrast, varying the distance-method window changes the set of candidate--muon pairs without changing the candidate selection.
\item \textbf{Distance-template fit range and histogram binning.}
The nominal distance-template configuration uses the fit range
$5\leq d\leq25~\mathrm{m}$ and 100 histogram bins. The lower boundary is
varied among 4, 5, and $6~\mathrm{m}$, the upper boundary among 20, 25,
and $30~\mathrm{m}$, and the number of bins among 80, 100, and 120. One
parameter is varied at a time while the remaining settings are kept fixed.
These variations quantify the dependence of the result on the fit range
and histogram binning.
\item \textbf{Isotope composition.}
The relative isotope composition is parameterized by the ratio
$N(^{8}\mathrm{He})/N(^{9}\mathrm{Li})$. The nominal analysis fixes this
ratio to $0.047$, as determined from the MC-truth composition.
Alternative ratios of $0$, $0.0143$, $0.03$, and $0.05$ are tested, with all other analysis settings kept fixed. The value $0.0143$ is taken from Ref.~\cite{DayaBay:2024xye}.
\end{itemize}

The relative systematic uncertainties obtained at the three event-selection
stages are summarized in Table~\ref{tab:systematic_budget}. Each uncertainty
is normalized to the nominal distance-constrained \lihe event contribution at
the corresponding selection stage. Since the nominal and varied configurations
are evaluated using the same finite simulation sample, their statistical
fluctuations are largely correlated and therefore partially cancel in the
differences. Residual finite-sample effects may nevertheless remain, particularly after the vetoes, where the isotope statistics are limited. For the distance-template source, the uncertainties associated with the fit range and histogram binning are evaluated separately and then combined in quadrature. The quoted values therefore provide conservative estimates of the dependence on the analysis configuration.








\begin{table*}[!t]
    \caption{
Relative systematic uncertainties in the \lihe event contribution obtained
with the DCJ-MuCAT fit at the three event-selection stages.
Each uncertainty is determined from the largest absolute deviation among
the predefined variations relative to the nominal result.
}
    \label{tab:systematic_budget}

    \small
    \renewcommand{\arraystretch}{1.15}

    \begin{tabular*}{\textwidth}{
        @{\extracolsep{\fill}}
        l
        c
        c
        c
        @{}
    }
    \toprule
    & \multicolumn{3}{c}{Relative systematic uncertainty (\%)} \\
    \cmidrule(lr){2-4}
    Systematic uncertainty source
    & Basic event selection
    & After SPN veto
    & After SPN and muon-track vetoes \\
    \midrule

    Muon energy-loss category boundaries
    & 1.1
    & 8.1
    & 24.8 \\

    Muon-association time windows
    & 0.8
    & 11.3
    & 34.5 \\

    Distance-template fit range and binning
    & 1.7
    & 15.6
    & 52.0 \\

    Isotope composition
    & 0.3
    & 0.1
    & 0.6 \\

    \midrule
    Total
    & 2.2
    & 20.9
    & 67.2 \\
    \bottomrule
    \end{tabular*}
\end{table*}

The distance-template fit range and histogram binning give the largest contribution at all three event-selection stages. The relative importance of the remaining sources varies with the selection. For the
basic event selection, the next-largest contribution comes from the muon
energy-loss category boundaries. After the spallation-neutron veto, the
muon-association time windows and muon energy-loss category boundaries
also give substantial contributions. The same pattern is observed after
the combined spallation-neutron and muon-track vetoes, with the
muon-association time windows giving the second-largest contribution. The
isotope-composition uncertainty remains small at all three stages. The
large relative uncertainties after the combined veto also reflect the
small nominal residual isotope contribution used in the normalization.

Changing the category boundaries redistributes the muons among the
energy-loss categories and modifies the category-specific muon rates. For
category $k$, the exponential slopes of the uncorrelated and isotope time
distributions are $R_{\mu,k}$ and $R_{\mu,k}+1/\tau_i$, respectively. As $R_{\mu,k}$ increases, the relative difference between these slopes decreases, making the two components more difficult to separate. The
extracted \lihe event contribution therefore becomes more sensitive to the
adopted categorization, particularly after the vetoes, where the remaining
isotope statistics are limited.

The muon-association-window variations quantify the sensitivity to the choice
of the candidate--muon association interval. For the J-MuCAT fit, the window defines the look-back interval used to identify the most
recent preceding muon in each category and therefore affects the accepted
candidate sample. For the TraDiTS method, extending the window includes more preceding muons and therefore more candidate--muon pairs in the distance distribution. A longer
distance-association window consequently increases the contribution from
unrelated muons without changing the candidate selection. As shown in Table~\ref{tab:systematic_budget}, the relatively large uncertainty from the association-window variations after the vetoes is driven mainly by the distance-association window. At this stage, the \lihe event contribution is reduced and the remaining sample is dominated by uncorrelated candidates.

For the distance-template variations, the basic-selection result is most sensitive to the lower boundary of the fit range, which determines how close the normalization region extends to the near-track region. After the vetoes, the upper boundary becomes more important because the event statistics decrease rapidly at large distances. Extending the fit range therefore includes low-statistics bins with larger fluctuations. These fluctuations can affect the fitted normalization of the uncorrelated template and consequently the inferred isotope contribution in the near-track region. This effect is amplified after the vetoes because the nominal residual isotope contribution is small. The histogram binning has a smaller effect than the fit-range variations at all three selection stages.

\subsection{Robustness Checks}
\label{sec:robustness_checks}
Additional checks are performed to assess the reliability of the statistical uncertainty. The dependence of the analysis on the fit formulation and track reconstruction is also examined. These checks are reported below
and are not included in the adopted systematic uncertainty budget.

As an implementation check, an Asimov dataset constructed from the
nominal four-category joint time model was fitted with the candidate-level
extended unbinned likelihood. The injected isotope contribution was recovered to numerical precision, confirming the normalization and implementation of the time-likelihood model.

\begin{itemize}
    \item \textbf{Bootstrap validation.}
A candidate-level bootstrap study is performed to assess the statistical
uncertainty reported by the DCJ-MuCAT fit. Candidate events
are resampled with replacement. For each bootstrap replica, the same resampled candidate indices are used to reconstruct the category-specific time inputs and the distance distribution. The distance constraint is then recalculated from the resampled distance distribution. This procedure preserves the statistical correlation between the time likelihood and the distance constraint due to their common candidate sample. The complete DCJ-MuCAT fit is then repeated
for each replica.

The resulting distribution of the fitted \lihe event contribution for the
basic selection is shown in Fig.~\ref{fig:bootstrap_validation}. The bootstrap validation is performed for the basic selection, where the isotope statistics are sufficient for a stable comparison. The bootstrap
distribution has a mean of 1978.3 events and a standard deviation of 65.1
events, consistent with the nominal fit result of
$1973.2 \pm 63.4$ events. 
The bootstrap distribution shows no significant displacement or strong asymmetry. The agreement between the bootstrap width and the nominal statistical uncertainty supports the reliability of the uncertainty estimate.

\begin{figure}[htbp]
    \centering
    \includegraphics[width=1\columnwidth]{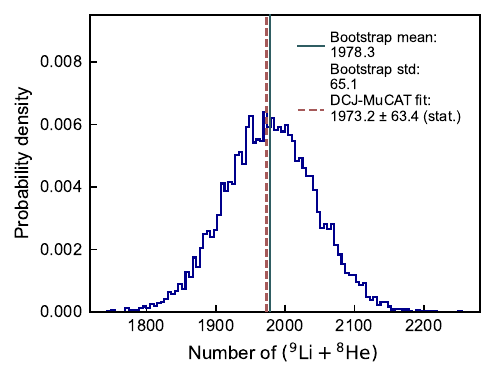}
    \caption{
Distribution of the fitted \lihe event contribution obtained from
$10\,000$ bootstrap samples of the DCJ-MuCAT fit for the basic selection.
The nominal fit result and its statistical uncertainty are shown for comparison.
}
    \label{fig:bootstrap_validation}
\end{figure}
\begin{table*}[!t]
    \caption{Results of the robustness tests for the J-MuCAT and
    DCJ-MuCAT fits. The values show the relative change in the fitted
    \lihe event contribution with respect to the nominal result at each
    event-selection stage.}
    \label{tab:robustness_studies}

    \small
    \renewcommand{\arraystretch}{1.15}

    \begin{tabular*}{\textwidth}{
        @{\extracolsep{\fill}}
        p{0.28\textwidth}
        l
        l
        l
        l
        @{}
    }
    \toprule
    & &
    \multicolumn{3}{c}{Relative change from nominal result (\%)} \\
    \cmidrule(lr){3-5}
    Robustness check
    & Method
    & Basic event selection
    & After SPN veto
    & After SPN and muon-track vetoes \\
    \midrule

    Track-direction smearing
    & DCJ-MuCAT
    & 6.4
    & 60.3
    & 100 \\

    Candidate--track distance smearing
    & DCJ-MuCAT
    & 0.7
    & 4.1
    & 68.5 \\

    One muon category
    & J-MuCAT
    & 4.9
    & 15.9
    & 86.6 \\

    Six muon categories
    & J-MuCAT
    & 1.0
    & 39.2
    & 144.5 \\

    \bottomrule
    \end{tabular*}
\end{table*}
\item \textbf{Muon-track smearing studies.}
The simulation provides truth-level muon tracks without a dedicated
track-reconstruction model. Possible reconstruction effects are assessed by applying parameterized smearings separately to the track direction and the candidate-to-track distance. The resulting changes are propagated
consistently to both the TraDiTS method and the muon-track-veto selection.

\textit{Track-direction smearing.}
Each truth-level track direction is perturbed by a Gaussian-distributed
angular deviation,
\begin{equation}
    \Delta\theta \sim \mathcal{N}(0,\sigma_\theta^2),
\end{equation}
where $\sigma_\theta$ denotes the standard deviation. The azimuthal
direction of the perturbation is sampled uniformly around the original
track direction. The track starting point and the magnitude of the
direction vector are kept unchanged. The values
$\sigma_\theta=1.0^\circ$, $1.5^\circ$, and $2.0^\circ$ are tested.
For each configuration, the candidate-to-track distances are recalculated
and the muon-track-veto selection is reapplied.

\textit{Candidate-to-track distance smearing.}
The calculated distance is modified according to
\begin{equation}
    d'=\max\left(0,\,d+\epsilon_d\right),
    \qquad
    \epsilon_d\sim\mathcal{N}(0,\sigma_d^2),
\end{equation}
where $\sigma_d=0.3$, $0.6$, and $0.9~\mathrm{m}$ are tested. The explicit
distance-bias parameter is fixed to zero. The smeared distances are used consistently in the muon-track-veto selection and the TraDiTS method. The resulting distance constraint is then propagated to the distance-constrained time fit.

For the basic selection and the spallation-neutron-veto selection, these
variations affect the track-distance estimate and therefore the distance
constraint, while leaving the candidate sample unchanged. After the combined spallation-neutron and muon-track vetoes, these variations also cause candidates to migrate across the track-veto boundary. This modifies the sample entering the fit. Across the tested configurations, the
distance-constrained \lihe event contribution after the combined veto
differs from the unsmeared result by up to $100.0\%$ for the track-direction
smearing and $68.5\%$ for the candidate-to-track distance smearing. For the largest tested track-direction smearing, $\sigma_\theta=2.0^\circ$, the fitted contribution reaches the physical boundary at zero. This corresponds to a relatively strong perturbation of the track direction. These large relative changes arise from the modified distance distribution and the migration of candidates across the veto boundary. Their impact is amplified by the small residual isotope contribution.

The tested smearing values are illustrative scenarios and do not correspond
to experimentally calibrated reconstruction resolutions. These studies are therefore treated as robustness checks rather than systematic uncertainty variations and are not included in the systematic uncertainty budget. The results are summarized in Table~\ref{tab:robustness_studies}.









\item \textbf{Dependence on muon categorization.} Muon categorization is introduced to account for differences in isotope production associated with muon energy loss. Separating muons into energy-loss categories also reduces the effective muon rate within each category, improving the temporal separation between the isotope and uncorrelated components.
The dependence of the categorized time fit on the number of muon categories is examined using one, two, five, and six categories. The nominal four-category configuration is used as the reference. The one-category configuration
corresponds to the conventional TSLM formulation, in which
all selected muons are treated as a single Poisson process. The one-category
result differs from the nominal result by $4.9\%$, $15.9\%$, and $86.6\%$
for the basic, SPN-veto, and combined SPN and muon-track-veto selections,
respectively. The corresponding six-category differences are $1.0\%$,
$39.2\%$, and $144.5\%$.

For the basic event selection, the fitted \lihe event contribution depends
only weakly on the number of categories. The dependence becomes substantially
stronger after the vetoes as the remaining isotope statistics decrease. In
particular, the one-category fit approaches the physical boundary at zero
after the combined veto, whereas the nominal four-category fit retains a
finite isotope contribution. The two-category result remains close to the
nominal result, while the five- and six-category fits give larger central
values. This comparison indicates that muon categorization improves the sensitivity of the time fit relative to the conventional one-category formulation under the tested conditions. However, the fit becomes more sensitive to the number of categories when the residual isotope statistics are limited.

These results illustrate the competing effects of the categorization: too few categories leave high category-specific muon rates and weak temporal separation, while too many categories distribute the limited isotope statistics among more fit components. Since only a limited set of discrete category schemes is examined, this study is treated as a fit-formulation robustness check rather than an additional systematic uncertainty. The representative
results are summarized in Table~\ref{tab:robustness_studies}.
\end{itemize}

\section{Conclusion}

This work develops complementary time- and distance-based methods for
estimating residual cosmogenic \lihe backgrounds in large neutrino
detectors.
The conventional TSLM method is extended to an event-level joint likelihood with muon energy-loss categories and corresponding muon rates.
A complementary spatial estimate is obtained using the TraDiTS method and incorporated as a Gaussian constraint on the total isotope contribution in the J-MuCAT fit.
The methods are evaluated in simulation after the basic selection, the spallation-neutron veto, and the combined spallation-neutron and muon-track vetoes.
Across these selections, the J-MuCAT fit gives smaller statistical uncertainties than the TSLM fit, while the combination of time and distance information remains effective when the time-only extraction is weakened by limited isotope statistics.
For the event selections studied in this work, the DCJ-MuCAT fit remains
consistent with the simulation-truth values and reduces the statistical
uncertainty by approximately $43$--$48\%$ relative to the J-MuCAT fit and by approximately $15$--$18\%$ relative to the TraDiTS estimate.
The applicability studies show that the timing method performs over a broad range of muon rates and isotope fractions, with its performance depending on the category definition, $R_{\mu,k}\tau_{\rm Li}$, and the available muon-rate information.
The TraDiTS estimate remains nearly unbiased over the scanned range.
Constraining the category-dependent muon rates further improves the J-MuCAT fit, indicating that such constraints can be incorporated in real-data analyses when independent muon-rate measurements are available.

The systematic studies show that the DCJ-MuCAT result is most sensitive to the distance-template configuration and, after strong vetoes, to the muon-association and categorization choices.
The paired-bootstrap, Asimov, and track-smearing studies further test the statistical consistency, likelihood implementation, and reconstruction robustness of the methods.
The track-smearing tests indicate increased sensitivity after the combined veto, particularly under strong perturbations of the reconstructed muon tracks.

The methods are not tied to a specific detector geometry or analysis
configuration.
With detector-specific treatments of the muon rate, track reconstruction, event selection, and uncorrelated distance distribution, the same strategy can be adapted to other large liquid-scintillator detectors and potentially to water-based neutrino experiments.
Beyond estimating residual cosmogenic backgrounds, the extracted isotope contributions can also provide a basis for measurements of cosmogenic isotope production yields when combined with the corresponding muon exposure and detector normalization.
The approach therefore has broader value for both background control and studies of cosmogenic isotope production in future low-background neutrino measurements.

\section*{Acknowledgments}
We thank Yaoguang Wang for his contribution to  the generation of the simulation samples used in this work.
We thank Tao Huang for helpful discussions.
This work is supported in part by the National Natural Science Foundation of China under Grants No.~12342502 and No.~12125506, and by the National Key Research and Development Program of China under Grant No.~2024YFE0110503.







\bibliography{biblio}

\begin{thebibliography}{22}%
\makeatletter
\providecommand \@ifxundefined [1]{%
 \@ifx{#1\undefined}
}%
\providecommand \@ifnum [1]{%
 \ifnum #1\expandafter \@firstoftwo
 \else \expandafter \@secondoftwo
 \fi
}%
\providecommand \@ifx [1]{%
 \ifx #1\expandafter \@firstoftwo
 \else \expandafter \@secondoftwo
 \fi
}%
\providecommand \natexlab [1]{#1}%
\providecommand \enquote  [1]{``#1''}%
\providecommand \bibnamefont  [1]{#1}%
\providecommand \bibfnamefont [1]{#1}%
\providecommand \citenamefont [1]{#1}%
\providecommand \href@noop [0]{\@secondoftwo}%
\providecommand \href [0]{\begingroup \@sanitize@url \@href}%
\providecommand \@href[1]{\@@startlink{#1}\@@href}%
\providecommand \@@href[1]{\endgroup#1\@@endlink}%
\providecommand \@sanitize@url [0]{\catcode `\\12\catcode `\$12\catcode
  `\&12\catcode `\#12\catcode `\^12\catcode `\_12\catcode `\%12\relax}%
\providecommand \@@startlink[1]{}%
\providecommand \@@endlink[0]{}%
\providecommand \url  [0]{\begingroup\@sanitize@url \@url }%
\providecommand \@url [1]{\endgroup\@href {#1}{\urlprefix }}%
\providecommand \urlprefix  [0]{URL }%
\providecommand \Eprint [0]{\href }%
\providecommand \doibase [0]{https://doi.org/}%
\providecommand \selectlanguage [0]{\@gobble}%
\providecommand \bibinfo  [0]{\@secondoftwo}%
\providecommand \bibfield  [0]{\@secondoftwo}%
\providecommand \translation [1]{[#1]}%
\providecommand \BibitemOpen [0]{}%
\providecommand \bibitemStop [0]{}%
\providecommand \bibitemNoStop [0]{.\EOS\space}%
\providecommand \EOS [0]{\spacefactor3000\relax}%
\providecommand \BibitemShut  [1]{\csname bibitem#1\endcsname}%
\let\auto@bib@innerbib\@empty
\bibitem [{\citenamefont {An}\ \emph {et~al.}(2012)\citenamefont {An},
  \citenamefont {Bai}, \citenamefont {Balantekin} \emph
  {et~al.}}]{DayaBay:2012fng}%
  \BibitemOpen
  \bibfield  {author} {\bibinfo {author} {\bibfnamefont {F.~P.}\ \bibnamefont
  {An}}, \bibinfo {author} {\bibfnamefont {J.~Z.}\ \bibnamefont {Bai}},
  \bibinfo {author} {\bibfnamefont {A.~B.}\ \bibnamefont {Balantekin}}, \emph
  {et~al.} (\bibinfo {collaboration} {Daya Bay Collaboration}),\ }\bibfield
  {title} {\bibinfo {title} {Observation of electron-antineutrino disappearance
  at {Daya Bay}},\ }\href@noop {} {\bibfield  {journal} {\bibinfo  {journal}
  {Phys. Rev. Lett.}\ }\textbf {\bibinfo {volume} {108}},\ \bibinfo {pages}
  {171803} (\bibinfo {year} {2012})}\BibitemShut {NoStop}%
\bibitem [{\citenamefont {de~Kerret}\ \emph {et~al.}(2020)\citenamefont
  {de~Kerret}, \citenamefont {Abrah{\~a}o}, \citenamefont {Almaz{\'a}n} \emph
  {et~al.}}]{DoubleChooz:2019qbj}%
  \BibitemOpen
  \bibfield  {author} {\bibinfo {author} {\bibfnamefont {H.}~\bibnamefont
  {de~Kerret}}, \bibinfo {author} {\bibfnamefont {T.}~\bibnamefont
  {Abrah{\~a}o}}, \bibinfo {author} {\bibfnamefont {H.}~\bibnamefont
  {Almaz{\'a}n}}, \emph {et~al.} (\bibinfo {collaboration} {Double Chooz
  Collaboration}),\ }\bibfield  {title} {\bibinfo {title} {Double chooz
  $\theta_{13}$ measurement via total neutron capture detection},\ }\href@noop
  {} {\bibfield  {journal} {\bibinfo  {journal} {Nat. Phys.}\ }\textbf
  {\bibinfo {volume} {16}},\ \bibinfo {pages} {558} (\bibinfo {year}
  {2020})}\BibitemShut {NoStop}%
\bibitem [{\citenamefont {Jeon}\ \emph {et~al.}(2025)\citenamefont {Jeon},
  \citenamefont {Kim}, \citenamefont {Choi} \emph {et~al.}}]{RENO:2024msr}%
  \BibitemOpen
  \bibfield  {author} {\bibinfo {author} {\bibfnamefont {S.}~\bibnamefont
  {Jeon}}, \bibinfo {author} {\bibfnamefont {H.~I.}\ \bibnamefont {Kim}},
  \bibinfo {author} {\bibfnamefont {J.~H.}\ \bibnamefont {Choi}}, \emph
  {et~al.} (\bibinfo {collaboration} {RENO Collaboration}),\ }\bibfield
  {title} {\bibinfo {title} {Measurement of reactor antineutrino oscillation
  parameters using the full 3800-day dataset of the {RENO} experiment},\
  }\href@noop {} {\bibfield  {journal} {\bibinfo  {journal} {Phys. Rev. D}\
  }\textbf {\bibinfo {volume} {111}},\ \bibinfo {pages} {112006} (\bibinfo
  {year} {2025})}\BibitemShut {NoStop}%
\bibitem [{\citenamefont {Abusleme}\ \emph
  {et~al.}(2026{\natexlab{a}})\citenamefont {Abusleme}, \citenamefont {Adam},
  \citenamefont {Adamowicz} \emph {et~al.}}]{JUNO:2025gmd}%
  \BibitemOpen
  \bibfield  {author} {\bibinfo {author} {\bibfnamefont {A.}~\bibnamefont
  {Abusleme}}, \bibinfo {author} {\bibfnamefont {T.}~\bibnamefont {Adam}},
  \bibinfo {author} {\bibfnamefont {K.}~\bibnamefont {Adamowicz}}, \emph
  {et~al.} (\bibinfo {collaboration} {JUNO Collaboration}),\ }\bibfield
  {title} {\bibinfo {title} {Measurement of reactor neutrino oscillation with
  the first {JUNO} data},\ }\href@noop {} {\bibfield  {journal} {\bibinfo
  {journal} {Nature}\ }\textbf {\bibinfo {volume} {654}},\ \bibinfo {pages}
  {343} (\bibinfo {year} {2026}{\natexlab{a}})}\BibitemShut {NoStop}%
\bibitem [{\citenamefont {Abusleme}\ \emph
  {et~al.}(2022{\natexlab{a}})\citenamefont {Abusleme}, \citenamefont {Adam},
  \citenamefont {Ahmad} \emph {et~al.}}]{JUNO:2022lpc}%
  \BibitemOpen
  \bibfield  {author} {\bibinfo {author} {\bibfnamefont {A.}~\bibnamefont
  {Abusleme}}, \bibinfo {author} {\bibfnamefont {T.}~\bibnamefont {Adam}},
  \bibinfo {author} {\bibfnamefont {S.}~\bibnamefont {Ahmad}}, \emph {et~al.}
  (\bibinfo {collaboration} {JUNO Collaboration}),\ }\bibfield  {title}
  {\bibinfo {title} {Prospects for detecting the diffuse supernova neutrino
  background with {JUNO}},\ }\href@noop {} {\bibfield  {journal} {\bibinfo
  {journal} {J. Cosmol. Astropart. Phys.}\ }\textbf {\bibinfo {volume} {10}},\
  \bibinfo {pages} {033}}\BibitemShut {NoStop}%
\bibitem [{\citenamefont {Agostini}\ \emph {et~al.}(2018)\citenamefont
  {Agostini}, \citenamefont {Altenm{\"u}ller}, \citenamefont {Appel} \emph
  {et~al.}}]{BOREXINO:2018ohr}%
  \BibitemOpen
  \bibfield  {author} {\bibinfo {author} {\bibfnamefont {M.}~\bibnamefont
  {Agostini}}, \bibinfo {author} {\bibfnamefont {K.}~\bibnamefont
  {Altenm{\"u}ller}}, \bibinfo {author} {\bibfnamefont {S.}~\bibnamefont
  {Appel}}, \emph {et~al.} (\bibinfo {collaboration} {Borexino
  Collaboration}),\ }\bibfield  {title} {\bibinfo {title} {Comprehensive
  measurement of $pp$-chain solar neutrinos},\ }\href@noop {} {\bibfield
  {journal} {\bibinfo  {journal} {Nature}\ }\textbf {\bibinfo {volume} {562}},\
  \bibinfo {pages} {505} (\bibinfo {year} {2018})}\BibitemShut {NoStop}%
\bibitem [{\citenamefont {Empl}\ and\ \citenamefont
  {Hungerford}(2014)}]{Empl:2014ona}%
  \BibitemOpen
  \bibfield  {author} {\bibinfo {author} {\bibfnamefont {A.}~\bibnamefont
  {Empl}}\ and\ \bibinfo {author} {\bibfnamefont {E.~V.}\ \bibnamefont
  {Hungerford}},\ }\href@noop {} {\bibinfo {title} {{A FLUKA Study of
  $\beta$-delayed Neutron Emission for the Ton-size DarkSide Dark Matter
  Detector}}} (\bibinfo {year} {2014}),\ \Eprint
  {https://arxiv.org/abs/1407.6628} {arXiv:1407.6628 [astro-ph.IM]}
  \BibitemShut {NoStop}%
\bibitem [{\citenamefont {P{\v e}{\v c}}\ \emph {et~al.}(2024)\citenamefont
  {P{\v e}{\v c}}, \citenamefont {Kudryavtsev}, \citenamefont {Ara{\'u}jo}
  \emph {et~al.}}]{Pec:2023yic}%
  \BibitemOpen
  \bibfield  {author} {\bibinfo {author} {\bibfnamefont {V.}~\bibnamefont {P{\v
  e}{\v c}}}, \bibinfo {author} {\bibfnamefont {V.~A.}\ \bibnamefont
  {Kudryavtsev}}, \bibinfo {author} {\bibfnamefont {H.~M.}\ \bibnamefont
  {Ara{\'u}jo}}, \emph {et~al.},\ }\bibfield  {title} {\bibinfo {title}
  {Muon-induced background in a next-generation dark matter experiment based on
  liquid xenon},\ }\href@noop {} {\bibfield  {journal} {\bibinfo  {journal}
  {Eur. Phys. J. C}\ }\textbf {\bibinfo {volume} {84}},\ \bibinfo {pages} {481}
  (\bibinfo {year} {2024})}\BibitemShut {NoStop}%
\bibitem [{\citenamefont {Abe}\ \emph {et~al.}(2010)\citenamefont {Abe},
  \citenamefont {Enomoto}, \citenamefont {Furuno} \emph
  {et~al.}}]{KamLAND:2009zwo}%
  \BibitemOpen
  \bibfield  {author} {\bibinfo {author} {\bibfnamefont {S.}~\bibnamefont
  {Abe}}, \bibinfo {author} {\bibfnamefont {S.}~\bibnamefont {Enomoto}},
  \bibinfo {author} {\bibfnamefont {K.}~\bibnamefont {Furuno}}, \emph {et~al.}
  (\bibinfo {collaboration} {KamLAND Collaboration}),\ }\bibfield  {title}
  {\bibinfo {title} {Production of radioactive isotopes through cosmic muon
  spallation in {KamLAND}},\ }\href@noop {} {\bibfield  {journal} {\bibinfo
  {journal} {Phys. Rev. C}\ }\textbf {\bibinfo {volume} {81}},\ \bibinfo
  {pages} {025807} (\bibinfo {year} {2010})}\BibitemShut {NoStop}%
\bibitem [{\citenamefont {de~Kerret}\ \emph {et~al.}(2018)\citenamefont
  {de~Kerret}, \citenamefont {Abrah{\~a}o}, \citenamefont {Almazan} \emph
  {et~al.}}]{DoubleChooz:2018kvj}%
  \BibitemOpen
  \bibfield  {author} {\bibinfo {author} {\bibfnamefont {H.}~\bibnamefont
  {de~Kerret}}, \bibinfo {author} {\bibfnamefont {T.}~\bibnamefont
  {Abrah{\~a}o}}, \bibinfo {author} {\bibfnamefont {H.}~\bibnamefont
  {Almazan}}, \emph {et~al.} (\bibinfo {collaboration} {Double Chooz
  Collaboration}),\ }\bibfield  {title} {\bibinfo {title} {Yields and
  production rates of cosmogenic {$^{9}$Li} and {$^{8}$He} measured with the
  {Double Chooz} near and far detectors},\ }\href@noop {} {\bibfield  {journal}
  {\bibinfo  {journal} {JHEP}\ }\textbf {\bibinfo {volume} {11}},\ \bibinfo
  {pages} {053}}\BibitemShut {NoStop}%
\bibitem [{\citenamefont {Hagner}\ \emph {et~al.}(2000)\citenamefont {Hagner},
  \citenamefont {von Hentig}, \citenamefont {Heisinger} \emph
  {et~al.}}]{Hagner:2000xb}%
  \BibitemOpen
  \bibfield  {author} {\bibinfo {author} {\bibfnamefont {T.}~\bibnamefont
  {Hagner}}, \bibinfo {author} {\bibfnamefont {R.}~\bibnamefont {von Hentig}},
  \bibinfo {author} {\bibfnamefont {B.}~\bibnamefont {Heisinger}}, \emph
  {et~al.},\ }\bibfield  {title} {\bibinfo {title} {Muon-induced production of
  radioactive isotopes in scintillation detectors},\ }\href@noop {} {\bibfield
  {journal} {\bibinfo  {journal} {Astropart. Phys.}\ }\textbf {\bibinfo
  {volume} {14}},\ \bibinfo {pages} {33} (\bibinfo {year} {2000})}\BibitemShut
  {NoStop}%
\bibitem [{\citenamefont {Bellini}\ \emph {et~al.}(2013)\citenamefont
  {Bellini}, \citenamefont {Benziger}, \citenamefont {Bick} \emph
  {et~al.}}]{Borexino:2013id}%
  \BibitemOpen
  \bibfield  {author} {\bibinfo {author} {\bibfnamefont {G.}~\bibnamefont
  {Bellini}}, \bibinfo {author} {\bibfnamefont {J.}~\bibnamefont {Benziger}},
  \bibinfo {author} {\bibfnamefont {D.}~\bibnamefont {Bick}}, \emph {et~al.}
  (\bibinfo {collaboration} {Borexino Collaboration}),\ }\bibfield  {title}
  {\bibinfo {title} {Cosmogenic backgrounds in {Borexino} at 3800~m
  water-equivalent depth},\ }\href@noop {} {\bibfield  {journal} {\bibinfo
  {journal} {J. Cosmol. Astropart. Phys.}\ }\textbf {\bibinfo {volume} {08}},\
  \bibinfo {pages} {049}}\BibitemShut {NoStop}%
\bibitem [{\citenamefont {Lee}\ \emph {et~al.}(2022)\citenamefont {Lee},
  \citenamefont {Choi}, \citenamefont {Jang} \emph {et~al.}}]{RENO:2022xbr}%
  \BibitemOpen
  \bibfield  {author} {\bibinfo {author} {\bibfnamefont {H.~G.}\ \bibnamefont
  {Lee}}, \bibinfo {author} {\bibfnamefont {J.~H.}\ \bibnamefont {Choi}},
  \bibinfo {author} {\bibfnamefont {H.~I.}\ \bibnamefont {Jang}}, \emph
  {et~al.} (\bibinfo {collaboration} {RENO Collaboration}),\ }\bibfield
  {title} {\bibinfo {title} {Measurement of cosmogenic {$^{9}$Li} and
  {$^{8}$He} production rates at {RENO}},\ }\href@noop {} {\bibfield  {journal}
  {\bibinfo  {journal} {Phys. Rev. D}\ }\textbf {\bibinfo {volume} {106}},\
  \bibinfo {pages} {012005} (\bibinfo {year} {2022})}\BibitemShut {NoStop}%
\bibitem [{\citenamefont {An}\ \emph {et~al.}(2024)\citenamefont {An},
  \citenamefont {Bai}, \citenamefont {Balantekin} \emph
  {et~al.}}]{DayaBay:2024xye}%
  \BibitemOpen
  \bibfield  {author} {\bibinfo {author} {\bibfnamefont {F.~P.}\ \bibnamefont
  {An}}, \bibinfo {author} {\bibfnamefont {W.~D.}\ \bibnamefont {Bai}},
  \bibinfo {author} {\bibfnamefont {A.~B.}\ \bibnamefont {Balantekin}}, \emph
  {et~al.} (\bibinfo {collaboration} {Daya Bay Collaboration}),\ }\bibfield
  {title} {\bibinfo {title} {First measurement of the yield of $^{8}$he
  isotopes produced in liquid scintillator by cosmic-ray muons at {Daya Bay}},\
  }\href@noop {} {\bibfield  {journal} {\bibinfo  {journal} {Phys. Rev. D}\
  }\textbf {\bibinfo {volume} {110}},\ \bibinfo {pages} {L011101} (\bibinfo
  {year} {2024})}\BibitemShut {NoStop}%
\bibitem [{\citenamefont {Wen}\ \emph {et~al.}(2006)\citenamefont {Wen},
  \citenamefont {Cao}, \citenamefont {Luk} \emph {et~al.}}]{Wen:2006hx}%
  \BibitemOpen
  \bibfield  {author} {\bibinfo {author} {\bibfnamefont {L.}~\bibnamefont
  {Wen}}, \bibinfo {author} {\bibfnamefont {J.}~\bibnamefont {Cao}}, \bibinfo
  {author} {\bibfnamefont {K.-B.}\ \bibnamefont {Luk}}, \emph {et~al.},\
  }\bibfield  {title} {\bibinfo {title} {Measuring cosmogenic {$^{9}$Li}
  background in a reactor neutrino experiment},\ }\href@noop {} {\bibfield
  {journal} {\bibinfo  {journal} {Nucl. Instrum. Meth. A}\ }\textbf {\bibinfo
  {volume} {564}},\ \bibinfo {pages} {471} (\bibinfo {year}
  {2006})}\BibitemShut {NoStop}%
\bibitem [{\citenamefont {An}\ \emph {et~al.}(2013)\citenamefont {An},
  \citenamefont {An}, \citenamefont {Bai} \emph {et~al.}}]{DayaBay:2012yjv}%
  \BibitemOpen
  \bibfield  {author} {\bibinfo {author} {\bibfnamefont {F.~P.}\ \bibnamefont
  {An}}, \bibinfo {author} {\bibfnamefont {Q.}~\bibnamefont {An}}, \bibinfo
  {author} {\bibfnamefont {J.~Z.}\ \bibnamefont {Bai}}, \emph {et~al.}
  (\bibinfo {collaboration} {Daya Bay Collaboration}),\ }\bibfield  {title}
  {\bibinfo {title} {Improved measurement of electron antineutrino
  disappearance at {Daya Bay}},\ }\href@noop {} {\bibfield  {journal} {\bibinfo
   {journal} {Chin. Phys. C}\ }\textbf {\bibinfo {volume} {37}},\ \bibinfo
  {pages} {011001} (\bibinfo {year} {2013})}\BibitemShut {NoStop}%
\bibitem [{\citenamefont {Abusleme}\ \emph
  {et~al.}(2022{\natexlab{b}})\citenamefont {Abusleme}, \citenamefont {Adam},
  \citenamefont {Ahmad} \emph {et~al.}}]{JUNO:2021vlw}%
  \BibitemOpen
  \bibfield  {author} {\bibinfo {author} {\bibfnamefont {A.}~\bibnamefont
  {Abusleme}}, \bibinfo {author} {\bibfnamefont {T.}~\bibnamefont {Adam}},
  \bibinfo {author} {\bibfnamefont {S.}~\bibnamefont {Ahmad}}, \emph {et~al.}
  (\bibinfo {collaboration} {JUNO Collaboration}),\ }\bibfield  {title}
  {\bibinfo {title} {Juno physics and detector},\ }\href@noop {} {\bibfield
  {journal} {\bibinfo  {journal} {Prog. Part. Nucl. Phys.}\ }\textbf {\bibinfo
  {volume} {123}},\ \bibinfo {pages} {103927} (\bibinfo {year}
  {2022}{\natexlab{b}})}\BibitemShut {NoStop}%
\bibitem [{\citenamefont {Abusleme}\ \emph {et~al.}(2025)\citenamefont
  {Abusleme}, \citenamefont {Adam}, \citenamefont {Ahmad} \emph
  {et~al.}}]{JUNO:2024jaw}%
  \BibitemOpen
  \bibfield  {author} {\bibinfo {author} {\bibfnamefont {A.}~\bibnamefont
  {Abusleme}}, \bibinfo {author} {\bibfnamefont {T.}~\bibnamefont {Adam}},
  \bibinfo {author} {\bibfnamefont {S.}~\bibnamefont {Ahmad}}, \emph {et~al.}
  (\bibinfo {collaboration} {JUNO Collaboration}),\ }\bibfield  {title}
  {\bibinfo {title} {Potential to identify neutrino mass ordering with reactor
  antineutrinos at {JUNO}},\ }\href@noop {} {\bibfield  {journal} {\bibinfo
  {journal} {Chin. Phys. C}\ }\textbf {\bibinfo {volume} {49}},\ \bibinfo
  {pages} {033104} (\bibinfo {year} {2025})}\BibitemShut {NoStop}%
\bibitem [{nnd()}]{nndc}%
  \BibitemOpen
  \href@noop {} {\bibinfo {title} {National nuclear data center}},\ \bibinfo
  {howpublished} {\url{https://www.nndc.bnl.gov/}},\ \bibinfo {note} {retrieved
  15 August 2026}\BibitemShut {NoStop}%
\bibitem [{\citenamefont {Abusleme}\ \emph
  {et~al.}(2026{\natexlab{b}})\citenamefont {Abusleme} \emph
  {et~al.}}]{JUNO:2025fpc}%
  \BibitemOpen
  \bibfield  {author} {\bibinfo {author} {\bibfnamefont {A.}~\bibnamefont
  {Abusleme}} \emph {et~al.} (\bibinfo {collaboration} {JUNO}),\ }\bibfield
  {title} {\bibinfo {title} {{Initial performance results of the JUNO
  detector}},\ }\href@noop {} {\bibfield  {journal} {\bibinfo  {journal} {Chin.
  Phys. C}\ }\textbf {\bibinfo {volume} {50}},\ \bibinfo {pages} {043001}
  (\bibinfo {year} {2026}{\natexlab{b}})}\BibitemShut {NoStop}%
\bibitem [{\citenamefont {Bellini}\ \emph {et~al.}(2012)\citenamefont {Bellini}
  \emph {et~al.}}]{Bellini:2012te}%
  \BibitemOpen
  \bibfield  {author} {\bibinfo {author} {\bibfnamefont {G.}~\bibnamefont
  {Bellini}} \emph {et~al.} (\bibinfo {collaboration} {Borexino}),\ }\bibfield
  {title} {\bibinfo {title} {{Cosmic-muon flux and annual modulation in
  Borexino at 3800 m water-equivalent depth}},\ }\href@noop {} {\bibfield
  {journal} {\bibinfo  {journal} {JCAP}\ }\textbf {\bibinfo {volume} {05}},\
  \bibinfo {pages} {015}}\BibitemShut {NoStop}%
\bibitem [{\citenamefont {Adamson}\ \emph {et~al.}(2010)\citenamefont {Adamson}
  \emph {et~al.}}]{MINOS:2009njg}%
  \BibitemOpen
  \bibfield  {author} {\bibinfo {author} {\bibfnamefont {P.}~\bibnamefont
  {Adamson}} \emph {et~al.} (\bibinfo {collaboration} {MINOS}),\ }\bibfield
  {title} {\bibinfo {title} {{Observation of muon intensity variations by
  season with the MINOS far detector}},\ }\href@noop {} {\bibfield  {journal}
  {\bibinfo  {journal} {Phys. Rev. D}\ }\textbf {\bibinfo {volume} {81}},\
  \bibinfo {pages} {012001} (\bibinfo {year} {2010})}\BibitemShut {NoStop}%
\end{thebibliography}%



\end{document}